\documentclass[a4paper,11pt]{article}
\usepackage{pos}

\def\beq{\begin{equation}}
\def\eeq{\end{equation}}
\def\beqa{\begin{eqnarray}}
\def\eeqa{\end{eqnarray}}

\title{New results for the 4-loop massive cusp anomalous dimension}

\author*{Nikolaos Kidonakis}

\affiliation{Department of Physics, Kennesaw State University,\\
Kennesaw, GA 30144, USA}

\emailAdd{nkidonak@kennesaw.edu}

\abstract{I present new results for the massive cusp anomalous dimension at four loops in QCD. I combine partial exact and conjectured analytical expressions with an approximation that is based on the asymptotic behavior of the cusp anomalous dimension to produce the newest estimates. Detailed comparisons are made with previous approximate results that highlight the robustness of the calculational approach.}

\FullConference{43rd International Conference on High Energy Physics (ICHEP 2026)\\
30 July to 5 August , 2026\\
Natal, Brazil\\}

\begin{document}
\maketitle

\section{Introduction}

The cusp anomalous dimension, relevant to the cusp formed by two Wilson lines, controls the infrared behavior of perturbative QCD scattering amplitudes \cite{cusp1,cusp2,NK2loop,cusp3,NK3loop,AGG,NK4loop}. It is the simplest soft anomalous dimension in QCD, and it is an essential ingredient in calculations of soft anomalous dimensions for processes with more complicated color structures involving heavy quarks \cite{NKGS,NKtop,MFNK}.

Wilson (eikonal) lines are path ordered exponentials that describe the emission of soft gluons by partons. The path of the parton (in this case a massive quark) is a straight line in the direction of its four-velocity $v$. The Wilson lines can be represented by the expression
\beq                                                  
W({\lambda}_2,{\lambda}_1;x)=
P\exp\left(-ig\int_{{\lambda}_1}^{{\lambda}_2}d{\lambda}\; 
v{\cdot} A ({\lambda}v+x)\right) \, ,
\eeq
where $A$ is the gauge field and $P$ indicates path ordering.

The cusp angle formed by two eikonal lines is given by $\theta=\cosh^{-1}(v_1\cdot v_2/\sqrt{v_1^2 v_2^2})$. The speed of the massive quark is $\beta=\tanh(\theta/2)$; inversely, the cusp angle is given in terms of the quark speed via $\theta=\ln[(1+\beta)/(1-\beta)]$. Thus, the infinite range for the cusp angle, $0 \le \theta < \infty$, corresponds to the range $ 0 \le \beta < 1$ for the quark speed. The finite and small range of the latter will be crucial for the approximation that we will discuss. We use the small-$\beta$ ($\beta \to 0$) and large-$\beta$ ($\beta \to 1$) asymptotics of the cusp anomalous dimension to make an approximation \cite{NK2loop,NK3loop,NK4loop} that can be used at all orders. We prove the excellence of the approximation by comparing with known exact results at two and three loops, and we make predictions at four loops \cite{NK4loop}. Here, we provide two further approximations at four loops and show that they make almost no difference to the original one of Ref. \cite{NK4loop}, thus highlighting the robustness of our approach.

\section{Perturbative series for $\Gamma_{\rm cusp}$}

We write the perturbative series in the strong coupling for the cusp anomalous dimension as 
$\Gamma_{\rm cusp}=\sum_{n=1}^{\infty} (\alpha_s/\pi)^n \Gamma^{(n)}$. 
The cusp anomalous dimension at each order can be read off the coefficients of the ultraviolet poles of the corresponding eikonal loop diagrams.

The one-loop cusp anomalous dimension is given in terms of $\beta$ by
\beq
\Gamma^{(1)}=-C_F \left[\frac{(1+\beta^2)}{2\beta}\ln\left(\frac{1-\beta}{1+\beta}\right) +1 \right] \, .
\eeq

The two-loop cusp anomalous dimension is given by \cite{NK2loop,NK3loop,NK4loop}
\beqa
\Gamma^{(2)}&=&K_2 \, \Gamma^{(1)}+C_F C_A \left\{\frac{1}{2}+\frac{\zeta_2}{2}
+\frac{1}{2}\ln^2\left(\frac{1-\beta}{1+\beta}\right) \right. 
\nonumber \\ && 
{}+\frac{(1+\beta^2)}{4\beta}\left[\zeta_2\ln\left(\frac{1-\beta}{1+\beta}\right)-\ln^2\left(\frac{1-\beta}{1+\beta}\right) +\frac{1}{3}\ln^3\left(\frac{1-\beta}{1+\beta}\right)
-{\rm Li}_2\left(\frac{4\beta}{(1+\beta)^2}\right)\right] 
\nonumber \\ &&  
{}+\frac{(1+\beta^2)^2}{8\beta^2}\left[-\zeta_3-\zeta_2\ln\left(\frac{1-\beta}{1+\beta}\right)-\frac{1}{3}\ln^3\left(\frac{1-\beta}{1+\beta}\right) \right. 
\nonumber \\ && \hspace{23mm} \left. \left.
{}-\ln\left(\frac{1-\beta}{1+\beta}\right) {\rm Li}_2\left(\frac{(1-\beta)^2}{(1+\beta)^2}\right)
+{\rm Li}_3\left(\frac{(1-\beta)^2}{(1+\beta)^2}\right)\right] \right\}
\eeqa
where $K_2=C_A (67/36-\zeta_2/2)-5 n_f/18$.

The three-loop cusp anomalous dimension was derived in Ref. \cite{cusp3} and presented in terms of the cusp angle, and it has a more complicated expression. 
Further study and reexpressions in terms of $\beta$ were presented in \cite{NK3loop,NK4loop}. We can write the three-loop result as \cite{NK3loop,NK4loop}
\beqa
\Gamma^{(3)}&=& K_3 \Gamma^{(1)}
+2 K_2 \left(\Gamma^{(2)}-K_2 \Gamma^{(1)}\right) + C^{(3)}
\nonumber
\eeqa
where $K_3$ and $C^{(3)}$ have long expressions.

\section{Large-$\beta$ asymptotics of $\Gamma_{\rm cusp}$}

The large-$\beta$ behavior of $\Gamma_{\rm cusp}$ at $n$ loops is given by
\beq
\lim_{\beta \to 1} \Gamma^{(n)}= K_n  \lim_{\beta \to 1} \Gamma^{(1)}+ P_n = -C_F K_n \lim_{\beta \to 1} \ln\left(\frac{1-\beta}{2}\right)+R_n
\eeq
where $R_n=P_n-C_F K_n$, and the constants $P_n$ at 
one, two, and three loops are given by 
$P_1=0$, $P_2=(1/2) C_F C_A (1-\zeta_3)$,
\beq
P_3=K_2 C_F C_A (1-\zeta_3)
+C_F C_A^2 \left(-\frac{1}{2}+\frac{3}{4}\zeta_2-\frac{\zeta_3}{4}
+\frac{9}{8} \zeta_5-\frac{3}{4}\zeta_2 \zeta_3 \right) \, .
\eeq

\section{Small-$\beta$ asymptotics of $\Gamma_{\rm cusp}$}

We write the small-$\beta$ expansion of $\Gamma_{\rm cusp}$ as
$\Gamma^{(n)}=\Gamma^{(n)}_{\beta^2}+\Gamma^{(n)}_{\beta^4}+{\cal O}(\beta^6)$,
and we find at one loop
\beq
\Gamma^{(1)}_{\beta^2}=\frac{4}{3}C_F \beta^2 \, ,  \quad \quad \quad
\Gamma^{(1)}_{\beta^4}=\frac{8}{15}C_F \beta^4 \, ,
\eeq
and at two loops
\beq
\Gamma^{(2)}_{\beta^2}= \beta^2 \left[C_F C_A \left(\frac{94}{27}-\frac{4}{3}\zeta_2 \right)-\frac{20}{27} C_F n_f T_F \right] \, ,
\eeq
\beq
\Gamma^{(2)}_{\beta^4}= \beta^4 \left[C_F C_A \left(\frac{64}{45}-\frac{8}{15}\zeta_2 \right)-\frac{8}{27} C_F n_f T_F \right] \, .
\eeq
The longer expressions for the expansion at three loops are given in \cite{NK3loop,NK4loop}. The small-$\beta$ expansion at four loops was presented in \cite{NK4loop} based on the small-$\theta$ expansions in \cite{GLP} by using the relation $\theta=2\beta+(2/3) \beta^3+{\cal O}(\beta^5)$.

\section{Approximate expressions for $\Gamma_{\rm cusp}$ from asymptotics}

Approximate expressions for $\Gamma_{\rm cusp}$ from its asymptotics at small and large $\beta$ were derived at two loops in Ref. \cite{NK2loop}, at three loops in Ref. \cite{NK3loop}, and at four loops in Ref. \cite{NK4loop}.

To derive the approximate result, $\Gamma_A^{(n)}$ at $n$ loops, we start with the small-$\beta$ expansion of $\Gamma_{\rm cusp}$, with $\Gamma^{(n)}_{\beta^{2,4}}=\Gamma^{(n)}_{\beta^2}+\Gamma^{(n)}_{\beta^4}$, then add $K_n \Gamma^{(1)}$ and subtract its small-$\beta$ expansion. Thus, we get 
\beq
\Gamma^{(n)}_A=\Gamma^{(n)}_{\beta^{2,4}}-K_n \, \Gamma^{(1)}_{\beta^{2,4}}
+K_n \, \Gamma^{(1)} \, .
\label{Gammanapprox}
\eeq
The last two terms on the right in Eq. (\ref{Gammanapprox}) cancel against each other at small $\beta$ while
the first two terms on the right cancel against each other at large $\beta$.
As noted in the introduction, the approximation works because of the small range of $\beta$. It would not work if one were to use $\theta$ expansions due to the infinite range of $\theta$.

We note that at two loops, the $C_F n_f$ terms in $\Gamma_A^{(2)}$ are exact, but the $C_F C_A$ terms are not exact. At three loops, the $C_F^2 n_f$ and the $C_F n_f^2$ terms in $\Gamma^{(3)}_A$ are exact, but the $C_F C_A^2$ and $C_F C_A n_f$ terms are not exact. 

Explicit results for the cusp anomalous dimension can be calculated separately for each value of $n_f$. The value $n_f=3$ is relevant for charm quark production \cite{NKRV}, $n_f=4$ for bottom quark production \cite{NKRV}, and $n_f=5$ for top quark production \cite{NKtop}.

\begin{figure}[htbp]
\begin{center}
\includegraphics[width=93mm]{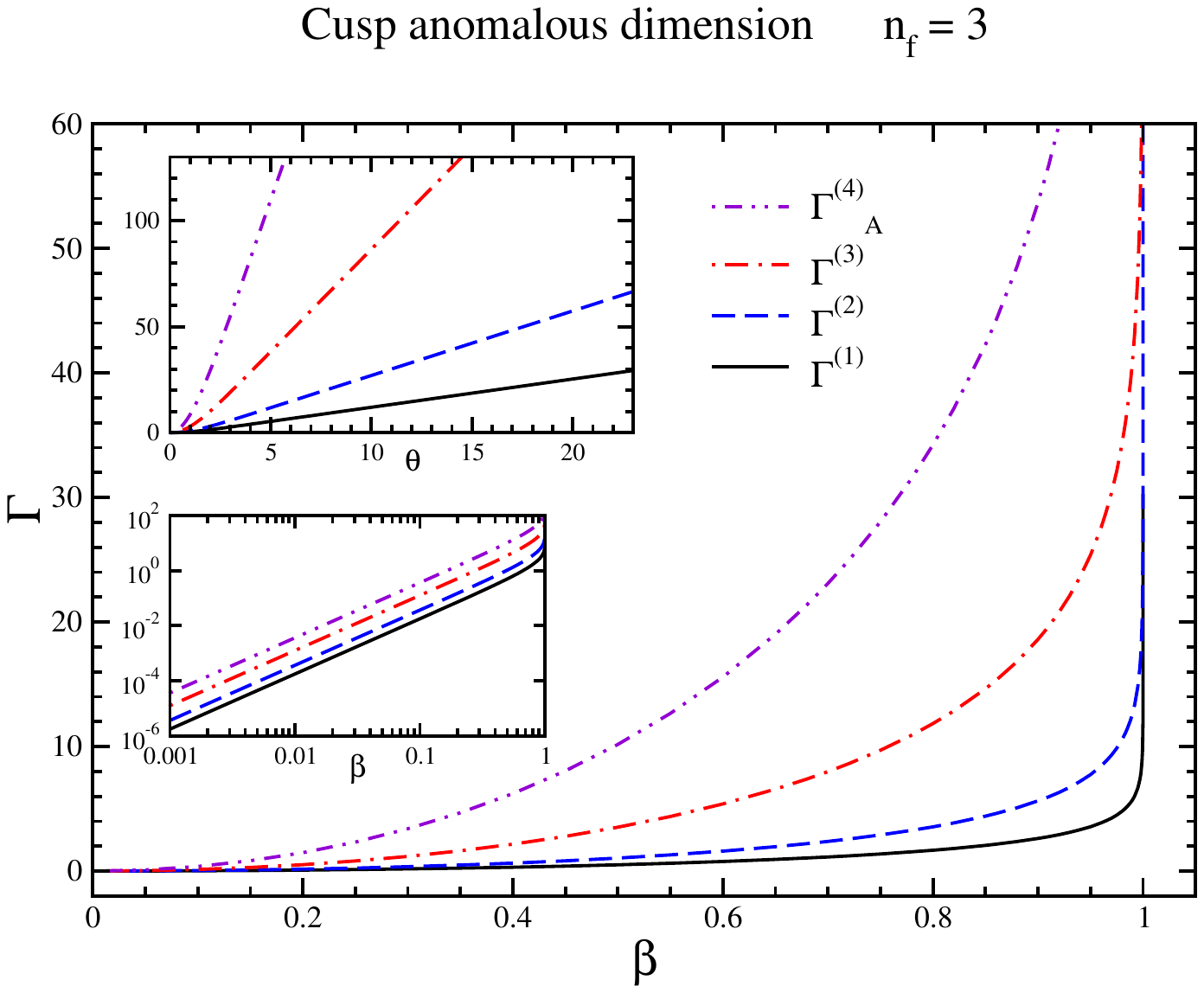}
\caption{The cusp anomalous dimension for $n_f=3$. Exact results are shown at one, two, and three loops, and approximate results are shown at four loops.}
\label{cuspnf3}
\end{center}
\end{figure}

In Fig. \ref{cuspnf3}, the exact results for the cusp anomalous dimension with $n_f=3$ at one, two, and three loops are displayed together with the approximate expression at four loops. The main plot shows the results as functions of $\beta$ in a linear scale, the bottom inset plot shows the same results in a logarithmic scale, and the top inset plot shows the results as functions of $\theta$ in a linear scale.
We note that $\beta=0$ corresponds to $\theta=0$, $\beta=0.01$ corresponds to $\theta \approx 0.02$, $\beta=0.5$ corresponds to $\theta \approx 1.1$, $\beta=0.99$ corresponds to $\theta \approx 5.3$, while $\beta=0.9999999999$ (the highest value we use) corresponds to $\theta \approx 23.7$.

\begin{figure}[htbp]
\begin{center}
\includegraphics[width=75mm]{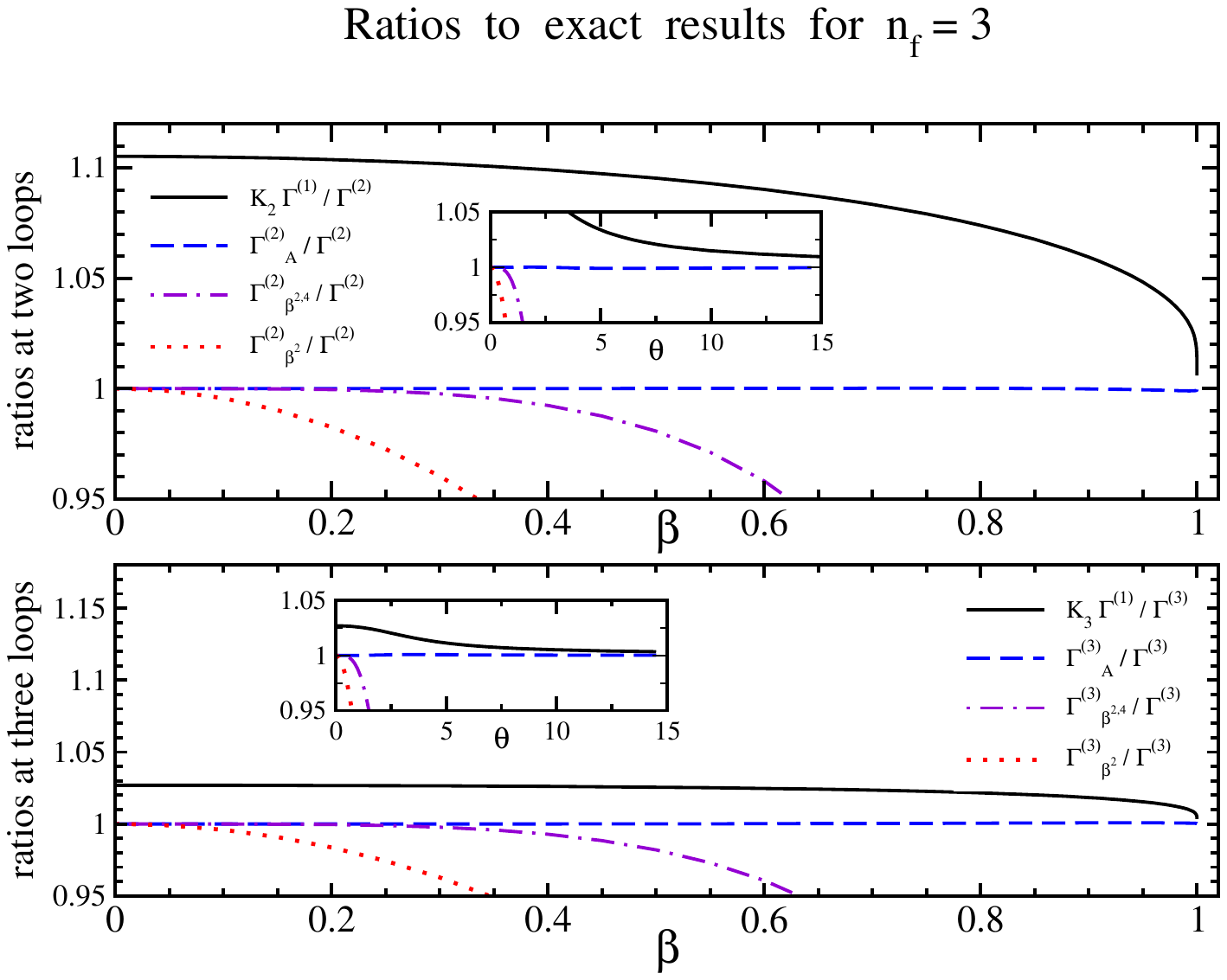}
\hspace{3mm}
\includegraphics[width=68mm]{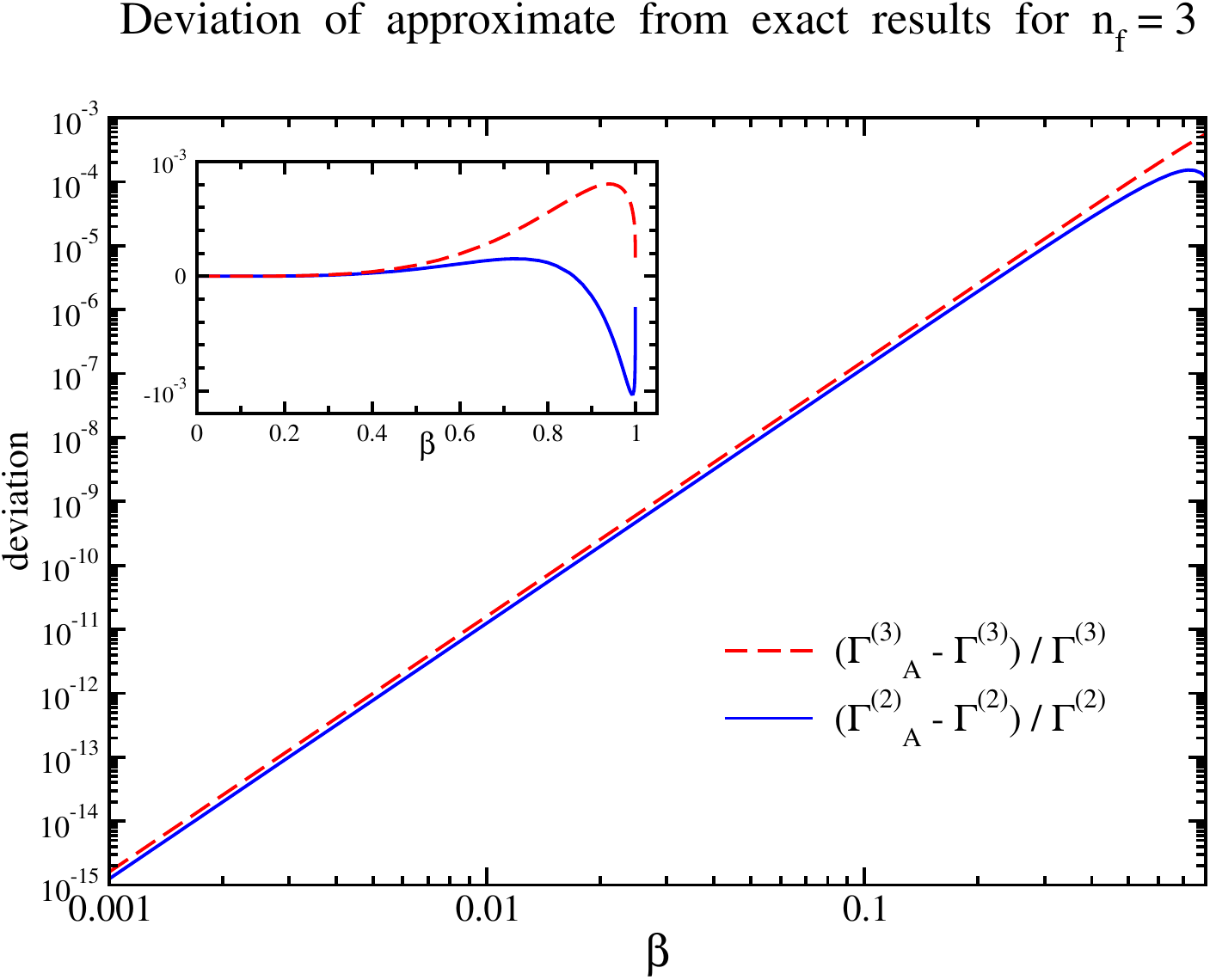}
\caption{Comparison of exact and approximate results at two and three loops for $n_f=3$.}
\label{Kgammanf3}
\end{center}
\end{figure}

In Fig. \ref{Kgammanf3}, we test the approximation by comparing the approximate results with the exact results at two and three loops for $n_f=3$. The two plots on the left show ratios of various quantities to the exact result for the cusp anomalous dimension at two loops (top plot) and three loops (bottom plot). We see that the small-$\beta$ expansions, through $\beta^2$ or through $\beta^4$, are only OK for small $\beta$, while $K_n \Gamma^{(1)}$ ($n=2$, 3) are only OK very close to $\beta \approx 1$. However, $\Gamma_A^{(2)}$ and $\Gamma_A^{(3)}$ are exceptionally good approximations to the two-loop and three-loop exact results, respectively, throughout the $\beta$ range; this is shown more clearly in the plot on the right where the deviations of the approximate from the exact results are shown at two and three loops. The approximations are astonishingly good, with e.g. deviations of ${\cal O}(10^{-15})$ at $\beta=0.001$, and they remain excellent throughout the $\beta$ range. The main plot shows the deviations for $\beta$ values from 0.001 to 0.8 in a logarithmic scale while the inset plot goes up to $\beta=0.9999999999$ in a linear scale. We note that the approximation is also exceptionally good separately for the individual color structures and also for other values of $n_f$.

\begin{figure}[htbp]
\begin{center}
\includegraphics[width=80mm]{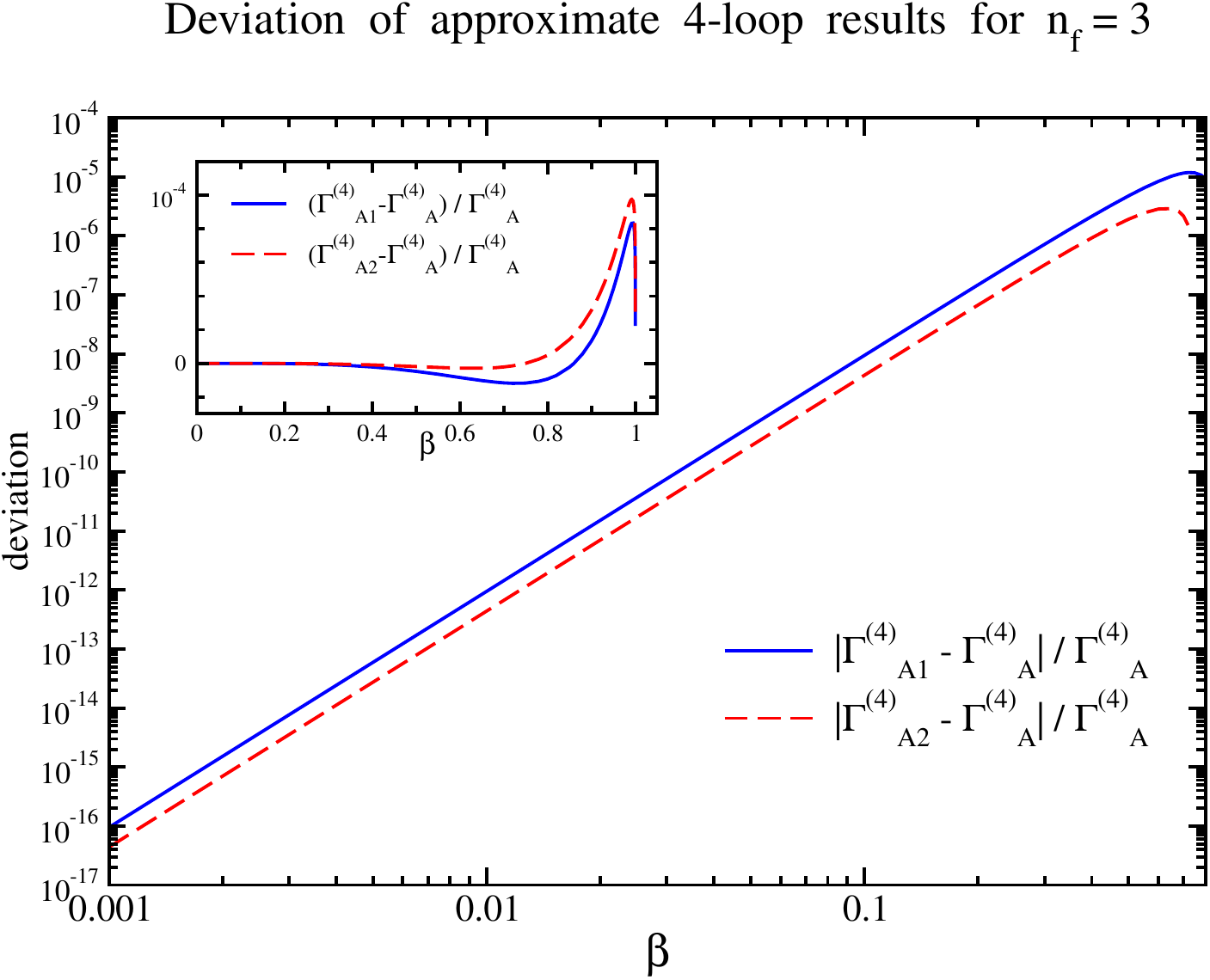}
\caption{Comparisons of three different approximations at four loops for $n_f=3$.}
\label{K4l}
\end{center}
\end{figure}

At four loops, the $C_F^3 n_f$, $C_F^2 n_f^2$, and $C_F n_f^3$ terms in $\Gamma^{(4)}_A$ are exact, but the $C_F C_A^3$, $C_F^2 C_A n_f$, $C_F C_A^2 n_f$, $C_F C_A n_f^2$, $d_F d_F$, and $d_F d_A$ terms are not exact. However, there exist conjectured exact results for the $C_F^2 C_A n_f$ terms and the $C_F C_A n_f^2$ terms \cite{cusp3,AGG}. Here, we use these results to construct a new alternate approximation which we denote as $\Gamma^{(4)}_{A1}$. Furthermore, we can also add the $\beta^6$ $d_F d_F$ terms \cite{BGHS} to $\Gamma^{(4)}_{A1}$ to construct yet another new approximation which we denote as $\Gamma^{(4)}_{A2}$.

In Fig. \ref{K4l}, we show the deviations of the new 4-loop approximations from the original one. We see that the original approximation is very robust since $\Gamma^{(4)}_{A1}$ and $\Gamma^{(4)}_{A2}$ are nearly identical to $\Gamma^{(4)}_A$. This shows that the 4-loop massive cusp anomalous dimension is now effectively determined numerically to very high precision.

\acknowledgments

This material is based upon work supported by the National Science Foundation under Grant No. PHY 2412071.

\end{document}